\documentclass[9pt,conference]{IEEEtran}
\usepackage{amssymb,amsthm,amsmath,array}
\usepackage{graphicx}
\usepackage{xspace}
\usepackage{stmaryrd}
\usepackage{xcolor}
\usepackage{mathtools}
\usepackage{float}
\usepackage{textcomp}
\usepackage{subfigure}
\usepackage{adjustbox}
\usepackage{optidef}
\usepackage{multicol}

\begin{document}
\title{Going Green in RAN Slicing}
\author{\IEEEauthorblockN{
        Hnin Pann Phyu\IEEEauthorrefmark{1}, 
         Razvan Stanica\IEEEauthorrefmark{2},
        Diala Naboulsi\IEEEauthorrefmark{3}, and
       Gwenael Poitau\IEEEauthorrefmark{4}
    }
    \IEEEauthorblockA{
        \IEEEauthorrefmark{1} Département de Génie Logiciel et TI,
École de Technologie Supérieure (ÉTS),
Montreal, Canada,
hnin.pann-phyu.1@ens.etsmtl\\
\IEEEauthorrefmark{2}Univ Lyon, INSA Lyon, Inria, CITI,
Villeurbanne, France,
razvan.stanica@insa-lyon.fr\\
\IEEEauthorrefmark{3} Département de Génie Logiciel et TI,
École de Technologie Supérieure (ÉTS),
Montreal, Canada,
diala.naboulsi@etsmtl.ca\\
\IEEEauthorrefmark{4}Dell Technologies, Ottawa, Canada, gwenael.poitau@dell.com
}
}

\maketitle


\begin{abstract}
Network slicing is essential for transforming future telecommunication networks into versatile service platforms, but it also presents challenges for sustainable network operations. While meeting the requirements of network slices incurs additional energy consumption compared to non-sliced networks, operators strive to offer diverse 5G and beyond services while maintaining energy efficiency. In this study, we address the issue of slice activation/deactivation to reduce energy consumption while maintaining the user quality of service (QoS). We employ Deep Contextual Multi-Armed Bandit and Thompson Sampling Contextual Multi-Armed Bandit agents to make activation/deactivation decisions for individual clusters. Evaluations are performed using the NetMob23 dataset, which captures the spatio-temporal consumption of various mobile services in France.  Our simulation results demonstrate that our proposed solutions provide significant reductions in network energy consumption while ensuring the QoS remains at a similar level compared to a scenario where all slice instances are active.
\end{abstract}

\section{Introduction}
The telecommunication industry accounts for approximately 2\% of total global carbon emissions~\cite{Cubukcuoglu2022}. Energy consumption will continue increasing in beyond 5G and 6G networks, where computationally intensive services will be largely deployed. In these future architectures, network slicing allows for the splitting of a physical network into multiple virtual networks, enabling mobile networks to cater to a diverse range of network services~\cite{10103689}.
Satisfying the requirements of different network slices, which includes ensuring performance isolation, comes at the cost of increased energy consumption. Meanwhile, we observe that the highest amount of energy is consumed in the radio access network (RAN), accounting approximately for 70\% of the overall network energy utilization~\cite{Piovesan2022}.

In this respect, several research works consider base station sleep schemes to further optimize the energy consumption in 5G networks~\cite{Han2013, Feng2017}. Applying these techniques directly in multi-services network slicing environments is more challenging, due to distinct temporal traffic patterns exhibited by different slice instances. Completely shutting down or putting the entire base station into sleep mode could significantly impact the quality of service (QoS) for users in particular slice instances. Moreover, several research works (\cite{Azimi2022, Rezazadeh2021,  Akin2022, Chergui2021}) consider optimizing the allocation of network slice resources (i.e. radio, CPU, transmission bandwidth and power) with respect to different network domains. However, none of these studies specifically addresses the handling of underutilized slice instances and the ability to deactivate them based on certain conditions.

This motivates us to propose a new approach: dynamically activating and deactivating slice instances based on their traffic patterns to enhance base station energy efficiency. However, deactivating certain slices to save energy might degrade the user QoS, while activating all slices at all times to maximize QoS results in significantly higher energy consumption. Therefore, the energy minimization objective shall be coupled with a QoS maximization objective~\cite{Chatzipapas2011}.

To manage the trade-off between the two objectives, we hereby introduce an EcoSlice, which is a slice instance using bare minimum resources and network functions. By that, it incurs lower energy consumption than typical slice instances. We consider the EcoSlice is up and running $24/7$ to provide a bare-minimum service. In some conditions, e.g., low traffic demand, operators may switch the users of other slices to this specific EcoSlice, without a significant QoS impact. All in all, in this work, we investigate the slice activation/deactivation problem with the aim of reducing the overall energy consumption while respecting as much as possible the QoS. 

\section{Methodology}
\subsection{System Model}
In a time-slotted system, we define $\tau$ as the slice activation/deactivation interval (SADI), during which slices remain continuously active or inactive. (De)activation decisions are made at the end of each $\tau$, for the subsequent $\tau+1$. 


Moreover, our system model is composed of: \emph{i)} \textbf{Set of slice instances.} This set comprises various virtual slices (e.g. eMBB, URLLC, mMTC, and EcoSlice) associated to base stations. Each slice instance is characterized by a specific Quality of Service Class Identifier (QCI) and its corresponding energy consumption. \emph{ii)} \textbf{Set of users:} This set represents the users associated with the slice instances deployed at base stations. Each user is defined by their required delay and traffic load demand. \emph{iii)} \textbf{Set of base stations:} This set includes all the base stations. Each base station has a set of possible slice activation/deactivation configurations.

We then define the overall energy consumption of a base station, which is composed of the energy consumption resulting from its associated slice instances (i.e. dynamic energy consumption) and a static energy consumption level. As mentioned, our objective is coupled with ensuring the user QoS. Hence, the user satisfaction is determined based on whether a user's delay requirement is satisfied or not. Finally, we use $\beta$ as a trade-off parameter between energy consumption and QoS: a larger $\beta$ gives more weight to the QoS.

\subsection{Proposed Solution}
\begin{figure}[ht!] 
    \centering
\includegraphics[scale=0.32]{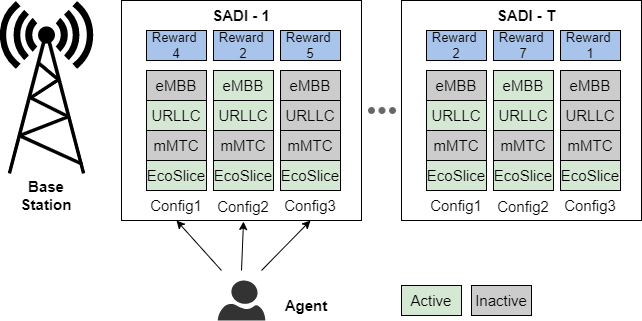}
    \caption{System architecture.}
    \label{dcmab}
\end{figure}

We propose two decentralized approaches: Deep Contextual Multi-Armed Bandit (DCMAB) and Thompson Sampling Contextual Multi-Armed Bandit (Thompson-C). The system architecture of our solutions is depicted in Figure~\ref{dcmab}. DCMAB and Thompson-C agents operate at the level of individual base stations. Each agent is presented with various configuration options of a base station, each associated with a reward. The agent's task is to select the most appropriate configuration at each time step ($\tau$) to achieve the overall objective. We model the problem of slice activation/deactivation as a contextual multi-armed bandit (MAB) problem.
\begin{itemize}
    \item \textbf{State}: The state space of an agent includes the overall energy consumption and average QoS of the base station.
    \item \textbf{Action}: An action of an agent is to choose which slice instances are active or not, and navigate users to and from an EcoSlice based on the state of their requested slice instance. 
    \item \textbf{Reward}: The reward function is designed to find the trade-off between energy consumption and QoS at each base station. 
\end{itemize}

\vspace{-12pt}

\section{Evaluation}
For the simulation, we assume slice instances are deployed on an application-basis, and we consider downlink traffic information of three distinct applications from the NetMob23 dataset: Facebook, Netflix, and Spotify, in the city of Orleans~\cite{netmob23}. Our agents operate at the individual base station level, which requires a pre-processing of the dataset. Specifically, we employ hierarchical clustering with the Ward linkage method to group the given tiles in Orleans based on their Pearson correlation values, and since the clustering results exhibit geographic continuity when plotted on the map, we consider each cluster to be equivalent to one base station. We train our models using data from 10 days (March 16-25, 2019).

\begin{figure}[htp]
  \centering
  \subfigure{\includegraphics[scale=0.45]{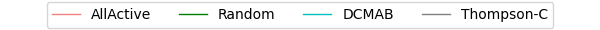}}
  \subfigure{\includegraphics[scale=0.22]{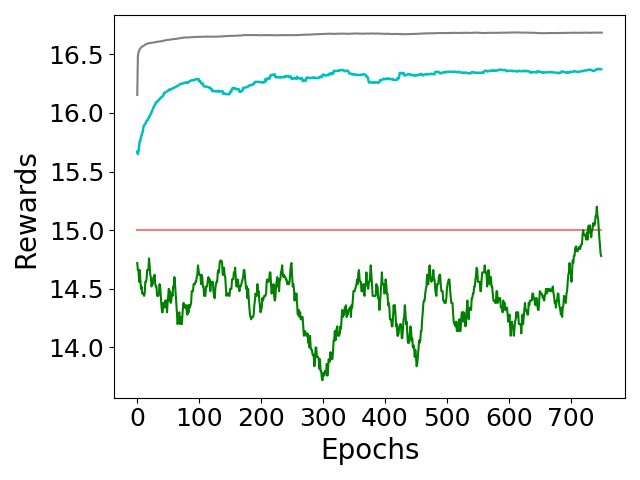}}\quad
  \subfigure{\includegraphics[scale=0.22]{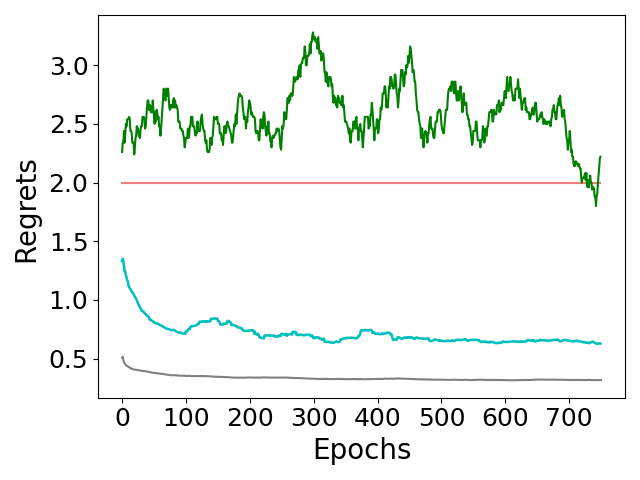}}
  \caption{Reward and regret obtained for $\beta=5$.}
  \label{rewardregret}
\end{figure}

We compare the performance of the proposed DCMAB and Thompson-C solutions with two counterparts: AllActive (all the slice instances are active) and Random. In Figure~\ref{rewardregret} (obtained for $\beta=5$), the DCMAB and Thompson-C agents exhibit better reward and regret trends than AllActive and Random strategies.  

\begin{figure}[htp]
  \centering
  \subfigure[Energy improvement over AllActive]{\includegraphics[scale=0.155]{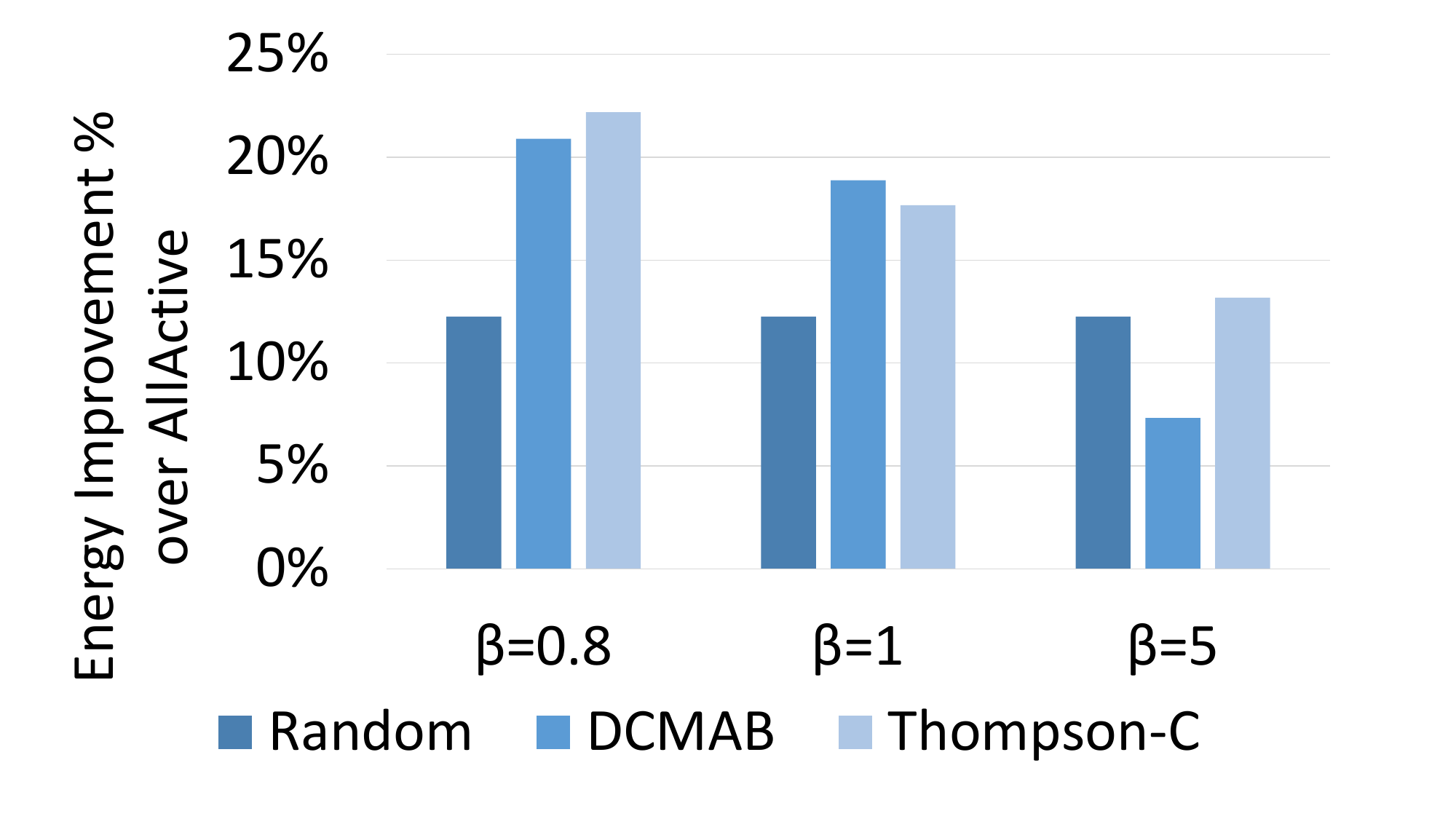}}\quad
  \subfigure[QoS]{\includegraphics[scale=0.155]{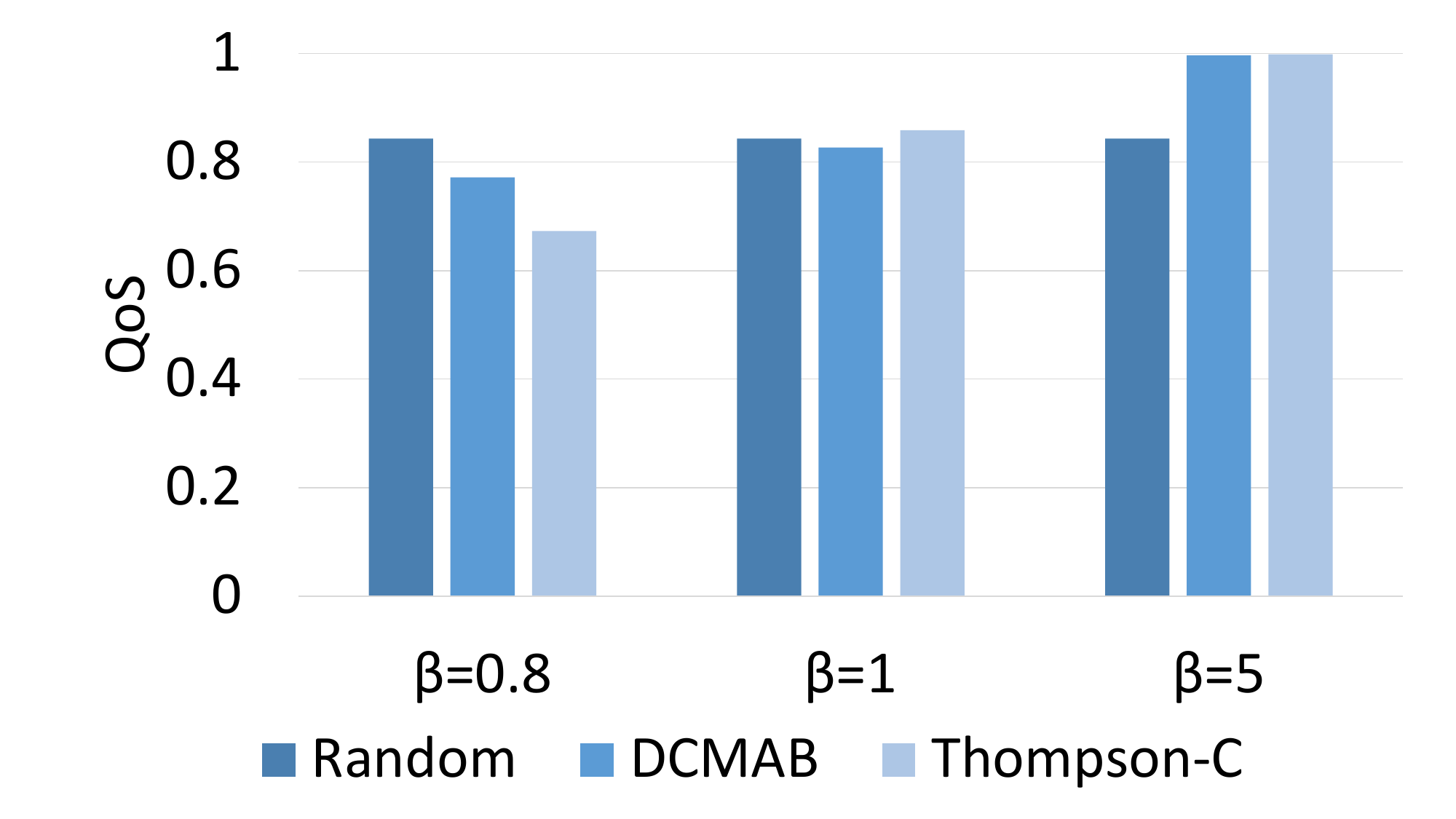}}
  \caption{Energy and QoS based on different $\beta$ values.}
  \label{energyqos}
\end{figure}

As further depicted in Figure~\ref{energyqos}, the energy improvement of DCMAB and Thompson-C compared with the current standard AllActive approach can reach 20\% for different $\beta$ values. The highest gain can be seen at $\beta=0.8$ but at the expense of a QoS degradation. It is important to note that reducing the energy consumption involves some compromise on QoS. However, as illustrated in Figure~\ref{energyqos} (b), when $\beta=5$, our agents ensure the same level of QoS as the AllActive solution, while providing significant energy gains. 

In all cases, Thompson-C outperforms DCMAB, but this comes with a much higher computational cost, as the execution time for Thompson-C is approximately 100 times larger. That being said, Thomson-C can be a favorable solution for a system with no computing time constraints, while DCMAB is better suited for real-time decision-making systems.

\section{Conclusion}

Our work focuses on enhancing energy efficiency in RAN slicing by addressing the slice activation/deactivation problem. We propose state-aware MAB approaches, namely DCMAB and Thompson-C, where an agent aims to activate the optimal slice instances while maintaining a given QoS level. Our results, based on the NetMob23 dataset, demonstrate that our proposed solutions significantly reduce energy consumption at the base station level while preserving QoS.

\bibliographystyle{IEEEtran}
\bibliography{references}
\end{document}